# Specification uncertainty:

# What the disruption index tells us about the (hidden) multiverse of bibliometric indicators

Christian Leibel[1,2]; Lutz Bornmann[1]

[1] Science Policy and Strategy Department

Administrative Headquarters of the Max Planck Society

Hofgartenstr. 8,

80539 Munich, Germany.

Email: christian.leibel.extern@gv.mpg.de, bornmann@gv.mpg.de

[2] Department of Sociology

Ludwig-Maximilians-Universität München

Konradstr. 6,

80801 Munich, Germany.

**Abstract**

Following Funk and Owen-Smith (2017), Wu et al. (2019) proposed the disruption index ($DI_1$) as a bibliometric indicator that measures disruptive and consolidating research. When we summarized the literature on the disruption index for our recently published review article (Leibel & Bornmann, 2024), we noticed that the calculation of disruption scores comes with numerous (hidden) degrees of freedom. In this Letter to the Editor, we explain based on the $DI_1$ (as an example) why the analytical flexibility of bibliometric indicators potentially endangers the credibility of research and advertise the application of multiverse-style methods to increase the transparency of the research.

When analyzing data, researchers have to make a multitude of decisions that affect the results of their research. Since there is often more than one justifiable approach to conducting data analysis, different (teams of) analysts may potentially arrive at different results when tasked with answering the same research question. Recent evidence shows the potential extent of uncertainty in empirical research: In a study by Schweinsberg et al. (2021), 19 analysts were given the same data and research question, but arrived at a broad range of results because of different analytical approaches and divergent operationalizations of key variables. The findings of Schweinsberg et al. (2021) and similar studies (Breznau et al., 2022; Huntington-Klein et al., 2021; Silberzahn et al., 2018) are highly relevant for indicator-based research like bibliometrics, because the specification of bibliometric indicators often comes with substantial degrees of freedom.[1] The possibility that different specifications of the same bibliometric indicator might lead to different research outcomes poses a potentially serious threat to the credibility of bibliometric research. In this Letter to the Editor, we use the disruption index ($DI_1$) as an example to illustrate the causes and consequences of the analytical flexibility of indicator-based bibliometric research and we encourage the application of multiverse-style methods to increase the transparency as well as the robustness of bibliometric analyses.

Shortly after Funk and Owen-Smith (2017) introduced the $DI_1$ as a measure of technological change[2], Wu et al. (2019) recognized its potential for the bibliometric study of transformative science. The $DI_1$ has started a new stream of research and plays a central role in no less than three *Nature* articles (Y. Lin et al., 2023; Park et al., 2023; Wu et al., 2019) and numerous other publications. When we summarized the literature on the $DI_1$ for our recently published review article (Leibel & Bornmann, 2024), we noticed that the calculation of disruption scores comes with numerous (hidden) degrees of freedom.

The $DI_1$ is closely related to measures of betweenness centrality (Freeman, 1977; Gebhart & Funk, 2023) and uses bibliographic coupling links to quantify historical discontinuities in the citation network of a focal paper (FP) (Leydesdorff & Bornmann, 2021). Bibliographic coupling links connect publications that cite the same references. The $DI_1$ ranges from -1 to 1 and is equivalent to the following ratio:

$$DI_1 = \frac{N_F - N_B}{N_F + N_B + N_R}$$

---

[1] While many-analyst studies deserve credit for shedding light on a "hidden universe of uncertainty" (Breznau et al., 2022), one should not brush over their limitations. A critical examination of current many-analyst studies shows that they probably paint "an overly pessimistic assessment of the credibility of science" (Auspurg & Brüderl, 2024)

[2] Funk and Owen-Smith (2017) use the name CD index, but we use disruption index ($DI_1$) in this Letter to the Editor and in our review article (Leibel & Bornmann, 2024).

$N_B$ is the number of citing papers that contain at least one bibliographic coupling link with the FP. These papers represent historical continuity because they connect the more recent literature with literature that predates the FP. Conversely, $N_F$ quantifies historical discontinuities by counting the number of papers that cite the FP without citing any of the FP's cited references. A large $N_F$ signals that the ideas that inspired the FP are no longer relevant for future research. Because $N_B$ is subtracted from $N_F$ in the numerator of the $DI_1$, positive disruption scores indicate that the FP "overshadows" (Liu et al., 2023) previous research. Negative disruption scores indicate that previous research still remains relevant after the publication of the FP. $N_R$ is the number of papers that cite the FP's cited references without citing the FP itself. Compared to $N_B$ and $N_F$, it is less clear what $N_R$ is supposed to represent. The purpose of $N_R$ may be to compare the citation impact of the FP to the citation impact of its cited references. $N_R$ reduces the disruption score of a FP considerably if the sum of the citations received by the cited references exceeds the citation count of the FP.

The calculation of the $DI_1$ hints at degrees of freedom that are often neglected in the literature. Since the justifiable modifications of the $DI_1$ are too numerous to discuss them all in this Letter to the Editor (there is an increasing multiverse of modifications), we refer the interested reader to our literature review (Leibel & Bornmann, 2024) where we provide an overview of the various alternatives to the $DI_1$ researchers have suggested so far. For the sake of brevity, we limit the illustration of the analytical flexibility of the $DI_1$ to three examples. First, Bornmann et al. (2020) point out that the $DI_1$ contains an implicit threshold $X$, such that a citing paper only counts towards $N_B$ if it cites at least $X$ of the FP's cited references. For the $DI_1$, $X = 1$, but one could just as well choose $X > 1$ with the argument that stronger bibliographic coupling links are a better indicator of historical continuity. Second, when calculating disruption scores it is common practice to use citation windows of $Y$ years. Third, the exclusion of FPs with less than $Z$ cited references (and/or citations) may help to avoid data artefacts. Table 1 explains the significance of $X$, $Y$, and $Z$ in detail and lists examples from the current literature.

If all combinations of $X$, $Y$, and $Z$ are justifiable, one would expect that all results support the same conclusion. However, as long as the set of possible results is not calculated and reported in its entirety, there is no way of knowing whether the results are consistent. In the case of the $DI_1$ and its variants, empirical evidence shows that the strength of the bibliographic coupling for $N_B$ (Bittmann et al., 2022; Bornmann et al., 2020; Bornmann & Tekles, 2021; Deng & Zeng, 2023; Wang et al., 2023), the length of the citation window (Bornmann & Tekles, 2019a; Liang et al., 2022), or the treatment of data artefacts (Holst et al., 2024; Liang et al., 2022; Ruan et al., 2021) can substantially affect research outcomes.

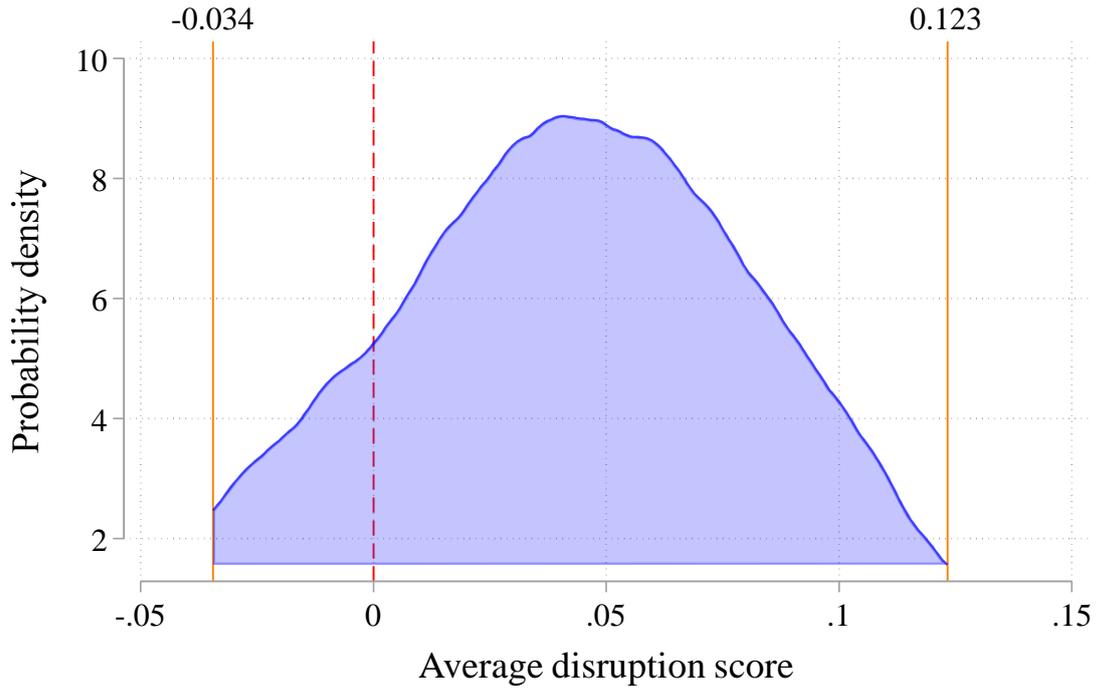

Figure 1: Kernel density distribution of the average disruption scores of 77 Nobel Prize winning papers published between 1985 and 2000 depending on the minimum number of bibliographic coupling links ($X$), the length of the citation window ($Y$), and the minimum number of cited references ($Z$).

For the purpose of illustration, we present the average disruption scores for a set of 77 Nobel Prize winning papers (NPs) published between 1985 and 2000. We collected the NPs from the dataset provided by Li et al. (2019) and used the Max Planck Society's inhouse version of the Web of Science (Clarivate) to calculate the disruption scores. We considered five choices for $X$ (1, 2, 3, 4, 5), three choices for $Y$ (3 years, 5 years, 10 years), and three choices for $Z$ (1 reference, 5 references, 10 references). Because each combination of $X$, $Y$, and $Z$ leads to a unique outcome, there is total of $5 \times 3 \times 3 = 45$ different results. Figure 1 shows both the range and the density distribution of the 45 results.[3] The majority of the average disruption scores (the average across the 77 NPs) are greater than zero, but some are smaller than zero. Even though the example is limited to $X$, $Y$, and $Z$, and does not consider additional factors that may affect disruption scores, it still unveils a broad range of possible results: The average disruption scores of the NPs range from -0.034 ($X = 1$, $Y = 3$, $Z = 1$) to 0.123 ($X = 5$, $Y = 10$, $Z = 1$).

---

[3] Figure 1 was created in Stata using the graphic scheme developed by Bischof (2017).

In light of the risk that research outcomes may vary greatly across unique combinations of $X$, $Y$, and $Z$ (as well as other factors), it seems problematic that in standard research practice researchers typically present analyses and results based on just one or maybe a few different specifications of the DI$_1$. Empirical results and their policy implications could hinge on arbitrarily chosen specifications of bibliometric indicators that are no more or less defensible than alternative specifications. If this uncertainty cannot be eliminated, it should at least be acknowledged. In empirical social research, it is standard practice to communicate and quantify the risk that samples may be unrepresentative in the form of standard errors, confidence intervals, and $p$ values. Similar, in bibliometrics, the credibility of research would profit from acknowledging that a result achieved with a specific variant of a bibliometric index may not be representative of the entire range of results that can be achieved with alternative indicator specifications.

We now turn to the methodological consequences of our observations. The widespread convention of presenting a main analysis and (maybe) a few robustness checks means that the reader of a bibliometric study gets to see results based on only a very limited number of indicator specifications. Sometimes, researchers may have good reasons for their selection of indicators. However, for want of convincing theoretical or statistical reasons to prefer any particular variant of an indicator to alternative specifications, analysts find themselves faced with a large set of justifiable indicators. In such a scenario, the important question is not "Which is the best indicator?" but rather "Which set of indicators deserves consideration?" (Young, 2018). This set – the *multiverse* of equally valid indicator specifications – "directly implies a multiverse of statistical results" (Steegen et al., 2016, p. 702), which should be reported in its entirety. This can be achieved with contemporary computational power, as was demonstrated by Muñoz and Young (2018). The authors show that widely used statistics software (like Stata) can both run and visualize several billion regressions.

Researchers from different disciplines have developed several multiverse-style methods like *multiverse analysis* (Steegen et al., 2016), *multimodel analysis* (Young & Holsteen, 2017), *specification-curve analysis* (Simonsohn et al., 2020), and *vibration of effects analysis* (Patel et al., 2015). All these methods build on the same core principle: Empirical results are not trustworthy if justifiable changes to the research strategy drastically alter the conclusions of a study. Multiverse-style methods can be thought of as very extensive and systematic robustness checks. Both conventional robustness checks and multiverse-style methods are guided by the notion that "a fragile inference is not worth taking seriously" (Leamer, 1985, p. 308). Only multiverse-style methods take this notion to its logical conclusion by transparently communicating the entire multiverse of statistical results. In the case of bibliometric indicators, this means that, ideally, all variants of an indicator that are supposed to capture the same concept (equally well) should lead to results that support the same conclusion.

We limit the focus of this Letter to the Editor to the multiverse of bibliometric indicators, but other aspects of the (bibliometric) research process like data collection (Harder, 2020), data processing (Steegen et al., 2016), and the specification of statistical models each come with their own multiverse. In empirical research, results are not only driven by raw data, but also by the decisions of the researchers that collect and analyze the data. The resulting uncertainty of research outcomes is, in the words of Chatfield (1995, p. 419), simply "a fact of life". It is not a unique feature of bibliometric indicators. We believe that multiverse-style methods may be of particular relevance for bibliometrics because bibliometric indicators, as we have exemplarily demonstrated for the $DI_1$, tend to have numerous variants. The variants create large multiverses that often remain unreported, e.g., due to the limitations of conventional robustness checks.

In this Letter to the Editor, we used the $DI_1$ as an example to illustrate the degrees of freedom that come with the specification of a bibliometric indicator. In a current research project, we are working on an empirical multiverse-style analysis to investigate the robustness of $DI_1$ scores. We believe that similar lines of argument may be applied to other bibliometric indicators like, e.g., interdisciplinarity indicators. Wang and Schneider (2020, p. 239) analyzed the consistency of interdisciplinarity measures and found "surprisingly deviant results when comparing measures that supposedly should capture similar features or dimensions of the concept of interdisciplinarity". Multiverse-style methods could be used to quantify the uncertainty of results in bibliometric interdisciplinarity research – as well as other streams of research – in order to find and eliminate sources of uncertainty. By unveiling the decision nodes required to calculate bibliometric indicators, multiverse-style methods could pave the way towards more robust indicator-based research.

Table 1: Examples of degrees of freedom in the calculation of the $DI_1$

| Aspect of specification | Examples in the literature | Explanation |
|---|---|---|
| Citing papers only count towards $N_B$ if they cite at least $X$ of the FP's cited references | $X = 1$: Funk and Owen-Smith (2017); Wu et al. (2019)<br>$X = 5$: Bornmann et al. (2020) | Since $N_B$ measures historical continuity via bibliographic coupling, $X$ determines how strong the bibliographic coupling between the FP and its citing paper needs to be in order to be interpreted as representing continuity. Citation data is noisy because citations occur for many different reasons. For example, both the FP and its citing papers may cite a highly cited reference for rhetorical reasons. Indeed, empirical evidence suggests that $DI_1$ scores are strongly driven by highly cited references (Liu et al., 2023). With $X > 1$, $N_B$ is less likely to be affected by noisy citation data. |
| Using citation windows of length $Y$ (in years) | Only publications published after FP<br>No citation window: Wei et al. (2023)<br>$Y = 3$ years: Liu et al. (2023)<br>$Y = 5$ years: Li and Chen (2022); Park et al. (2023)<br><br>All publications:<br>No citation window: Bittmann et al. (2022); Bornmann and Tekles (2019b, 2021); Z. Lin et al. (2023) | Citation windows address the fact that $DI_1$ scores depend on the time of measurement by taking a snapshot of every FP's citation network at a fixed point in time. It is a special feature of the $DI_1$ (and some of its variants) that citation windows may include publications that predate the FP if publications that only cite the FP's references were published prior to the FP. Thus, there are two versions of every citation window of length $Y$: One version that includes publications published after the FP and another version that includes all publications (even those published prior to the FP). Funk and Owen-Smith (2017) only include subsequent publications. |

Table 1: Examples of degrees of freedom in the calculation of the $DI_1$

| | | |
|---|---|---|
| Performing analyses only on FPs with at least $Z$ cited references (and/or citations) | No minimum: Park et al. (2023) $Z = 5$: Deng and Zeng (2023) $Z = 10$: Bornmann et al. (2020); Bornmann and Tekles (2019b); Sheng et al. (2023) | An FP with no cited references has $N_B = N_R = 0$ and is thus guaranteed to receive the maximum $DI_1$ score of 1 as long as it has at least one citation – a clear case of a data artefact. Such cases of $DI_1 = 1$ are highly misleading, particularly if the FP does actually cite references which are not covered by the respective literature database. The cited references are thus missing in the reconstructed citation network of the FP (Holst et al., 2024; Liang et al., 2022; Ruan et al., 2021). |

# Statements and Declarations

Competing interests: The authors do not declare any competing interests.

# Funding

This study uses ivsdb data. The ivsdb data are developed in cooperation between the Administrative Headquarters of the Max Planck Society (MPG) and the Information Retrieval Service for the institutes of the Chemical Physical Technical (IVS-CPT) Section of the MPG. The data stem from a bibliometrics database based on data provided by the German Kompetenznetzwerk Bibliometrie (KB, Competence Network Bibliometrics, funded by BMBF via grant 01PQ17001, see: http://www.bibliometrie.info/). The data are derived from the Science Citation Index - Expanded (SCI-E), the Social Sciences Citation Index (SSCI), the Conference Proceedings Citation Index - Science (CPCI-S), the Conference Proceedings Citation Index - Social Science & Humanities (CPCI-SSH), and the Arts and Humanities Citation Index (AHCI), provided by Clarivate and updated in calendar week 17 of 2024.

# References

Auspurg, K., & Brüderl, J. (2024). Toward a more credible assessment of the credibility of science by many-analyst studies. *Proceedings of the National Academy of Sciences, 121*(38), Article e2404035121. https://doi.org/10.1073/pnas.2404035121

Bischof, D. (2017). New graphic schemes for Stata: plotplain and plottig. *Stata Journal, 17*(3), 748-759. https://doi.org/10.1177/1536867x1701700313

Bittmann, F., Tekles, A., & Bornmann, L. (2022). Applied usage and performance of statistical matching in bibliometrics: The comparison of milestone and regular papers with multiple measurements of disruptiveness as an empirical example. *Quantitative Science Studies, 2*(4), 1246-1270. https://doi.org/10.1162/qss_a_00158

Bornmann, L., Devarakonda, S., Tekles, A., & Chacko, G. (2020). Are disruption index indicators convergently valid? The comparison of several indicator variants with assessments by peers. *Quantitative Science Studies, 1*(3), 1242-1259. https://doi.org/10.1162/qss_a_00068

Bornmann, L., & Tekles, A. (2019a). Disruption index depends on length of citation window. *Profesional De La Informacion, 28*(2), Article e280207. https://doi.org/10.3145/epi.2019.mar.07

Bornmann, L., & Tekles, A. (2019b). Disruptive papers published in *Scientometrics*. *Scientometrics, 120*(1), 331-336. https://doi.org/10.1007/s11192-019-03113-z

Bornmann, L., & Tekles, A. (2021). Convergent validity of several indicators measuring disruptiveness with milestone assignments to physics papers by experts. *Journal of Informetrics, 15*(3), Article 101159. https://doi.org/10.1016/j.joi.2021.101159

Breznau, N., Rinke, E. M., Wuttke, A., Nguyen, H. H. V., Adem, M., Adriaans, J., Alvarez-Benjumea, A., Andersen, H. K., Auer, D., Azevedo, F., Bahnsen, O., Balzer, D., Bauer, G., Bauer, P. C., Baumann, M., Baute, S., Benoit, V., Bernauer, J., Berning, C., . . . Zóltak, T. (2022). Observing many researchers using the same data and hypothesis reveals a hidden universe of uncertainty. *Proceedings of the National Academy of Sciences of the United States of America, 119*(44), Article e2203150119. https://doi.org/10.1073/pnas.2203150119

Chatfield, C. (1995). Model uncertainty, data mining and statistical inference. *Journal of the Royal Statistical Society. Series A (Statistics in Society), 158*(3), 419-466. https://doi.org/10.2307/2983440

Deng, N., & Zeng, A. (2023). Enhancing the robustness of the disruption metric against noise. *Scientometrics, 128*(4), 2419–2428. https://doi.org/10.1007/s11192-023-04644-2

Freeman, L. C. (1977). A set of measures of centrality based on betweenness. *Sociometry, 40*(1), 35-41. https://doi.org/10.2307/3033543

Funk, R. J., & Owen-Smith, J. (2017). A dynamic network measure of technological change. *Management Science, 63*(3), 791-817. https://doi.org/10.1287/mnsc.2015.2366

Gebhart, T., & Funk, R. J. (2023). *A mathematical framework for citation disruption*. arXiv. Retrieved September 11, 2023 from https://doi.org/10.48550/arXiv.2308.16363

Harder, J. A. (2020). The multiverse of methods: Extending the multiverse analysis to address data-collection decisions. *Perspectives on Psychological Science, 15*(5), 1158-1177. https://doi.org/10.1177/1745691620917678

Holst, V., Algaba, A., Tori, F., Wenmackers, S., & Ginis, V. (2024). *Dataset artefacts are the hidden drivers of the declining disruptiveness in science*. arXiv. Retrieved February 26, 2024 from https://doi.org/10.48550/arXiv.2402.14583

Huntington-Klein, N., Arenas, A., Beam, E., Bertoni, M., Bloem, J. R., Burli, P., Chen, N. B., Grieco, P., Ekpe, G., Pugatch, T., Saavedra, M., & Stopnitzky, Y. (2021). The influence of hidden researcher decisions in applied microeconomics. *Economic Inquiry, 59*(3), 944-960. https://doi.org/10.1111/ecin.12992

Leamer, E. E. (1985). Sensitivity analyses would help. *American Economic Review, 75*(3), 308-313. https://www.jstor.org/stable/1814801

Leibel, C., & Bornmann, L. (2024). What do we know about the disruption index in scientometrics? An overview of the literature. *Scientometrics, 129*(1), 601-639. https://doi.org/10.1007/s11192-023-04873-5

Leydesdorff, L., & Bornmann, L. (2021). Disruption indices and their calculation using web-of-science data: Indicators of historical developments or evolutionary dynamics? *Journal of Informetrics, 15*(4), Article 101219. https://doi.org/10.1016/j.joi.2021.101219

Li, J., & Chen, J. (2022). Measuring destabilization and consolidation in scientific knowledge evolution. *Scientometrics, 127*(10), 5819-5839. https://doi.org/10.1007/s11192-022-04479-3

Li, J., Yin, Y., Fortunato, S., & Wang, D. (2019). A dataset of publication records for Nobel laureates. *Scientific Data, 6*, Article 33. https://doi.org/10.1038/s41597-019-0033-6

Liang, G., Lou, Y., & Hou, H. (2022). Revisiting the disruptive index: Evidence from the Nobel Prize-winning articles. *Scientometrics, 127*(10), 5721-5730. https://doi.org/10.1007/s11192-022-04499-z

Lin, Y., Frey, C. B., & Wu, L. (2023). Remote collaboration fuses fewer breakthrough ideas. *Nature, 623*(7989), 987-991. https://doi.org/10.1038/s41586-023-06767-1

Lin, Z., Yin, Y., Liu, L., & Wang, D. (2023). SciSciNet: A large-scale open data lake for the science of science research. *Scientific Data, 10*(1), Article 315. https://doi.org/10.1038/s41597-023-02198-9

Liu, X., Zhang, C., & Li, J. (2023). Conceptual and technical work: Who will disrupt science? *Journal of Informetrics, 17*(3), Article 101432. https://doi.org/10.1016/j.joi.2023.101432

Muñoz, J., & Young, C. (2018). We ran 9 billion regressions: Eliminating false positives through computational model robustness. *Sociological Methodology, 48*(1), 1-33. https://doi.org/10.1177/0081175018777988

Park, M., Leahey, E., & Funk, R. J. (2023). Papers and patents are becoming less disruptive over time. *Nature, 613*(7942), 138-144. https://doi.org/10.1038/s41586-022-05543-x

Patel, C. J., Burford, B., & Ioannidis, J. P. A. (2015). Assessment of vibration of effects due to model specification can demonstrate the instability of observational associations. *Journal of Clinical Epidemiology, 68*(9), 1046-1058. https://doi.org/10.1016/j.jclinepi.2015.05.029

Ruan, X., Lyu, D., Gong, K., Cheng, Y., & Li, J. (2021). Rethinking the disruption index as a measure of scientific and technological advances. *Technological Forecasting and Social Change, 172*, Article 121071. https://doi.org/10.1016/j.techfore.2021.121071

Schweinsberg, M., Feldman, M., Staub, N., van den Akker, O. R., van Aert, R. C. M., van Assen, M., Liu, Y., Althoff, T., Heer, J., Kale, A., Mohamed, Z., Amireh, H., Prasad, V. V., Bernstein, A., Robinson, E., Snellman, K., Sommer, S. A., Otner, S. M. G., Robinson, D., . . . Uhlmann, E. L. (2021). Same data, different conclusions: Radical dispersion in empirical results when independent analysts operationalize and test the same hypothesis. *Organizational Behavior and Human Decision Processes, 165*, 228-249. https://doi.org/10.1016/j.obhdp.2021.02.003

Sheng, L., Lyu, D., Ruan, X., Shen, H., & Cheng, Y. (2023). The association between prior knowledge and the disruption of an article. *Scientometrics, 128*(8), 4731–4751. https://doi.org/10.1007/s11192-023-04751-0

Silberzahn, R., Uhlmann, E. L., Martin, D. P., Anselmi, P., Aust, F., Awtrey, E., Bahník, S., Bai, F., Bannard, C., Bonnier, E., Carlsson, R., Cheung, F., Christensen, G., Clay, R.,

Craig, M. A., Dalla Rosa, A., Dam, L., Evans, M. H., Cervantes, I. F., . . . Nosek, B. A. (2018). Many analysts, one data set: Making transparent how variations in analytic choices affect results. *Advances in Methods and Practices in Psychological Science, 1*(3), 337-356. https://doi.org/10.1177/2515245917747646

Simonsohn, U., Simmons, J. P., & Nelson, L. D. (2020). Specification curve analysis. *Nature Human Behaviour, 4*(11), 1208-1214. https://doi.org/10.1038/s41562-020-0912-z

Steegen, S., Tuerlinckx, F., Gelman, A., & Vanpaemel, W. (2016). Increasing transparency through a multiverse analysis. *Perspectives on Psychological Science, 11*(5), 702-712. https://doi.org/10.1177/1745691616658637

Wang, Q., & Schneider, J. W. (2020). Consistency and validity of interdisciplinarity measures. *Quantitative Science Studies, 1*(1), 239-263. https://doi.org/10.1162/qss_a_00011

Wang, S., Ma, Y., Mao, J., Bai, Y., Liang, Z., & Li, G. (2023). Quantifying scientific breakthroughs by a novel disruption indicator based on knowledge entities. *Journal of the Association for Information Science and Technology, 74*(2), 150-167. https://doi.org/10.1002/asi.24719

Wei, C., Li, J., & Shi, D. (2023). Quantifying revolutionary discoveries: Evidence from Nobel prize-winning papers. *Information Processing & Management, 60*(3), Article 103252. https://doi.org/10.1016/j.ipm.2022.103252

Wu, L., Wang, D., & Evans, J. A. (2019). Large teams develop and small teams disrupt science and technology. *Nature, 566*(7744), 378-382. https://doi.org/10.1038/s41586-019-0941-9

Young, C. (2018). Model uncertainty and the crisis in science. *Socius, 4*, 1-7. https://doi.org/10.1177/2378023117737206

Young, C., & Holsteen, K. (2017). Model uncertainty and robustness: A computational framework for multimodel analysis. *Sociological Methods & Research, 46*(1), 3-40. https://doi.org/10.1177/0049124115610347